\documentstyle[12pt,aps,pre,epsf,preprint,tighten]{revtex}

\begin{document}
\draft
\preprint{HKBU-CNS-9714}
\title{
Squeezed state dynamics of kicked quantum systems
}
\author{
Bambi Hu$^{[1,2]}$, Baowen Li$^{[1]}$, Jie Liu$^{[1,3]}$, and Ji-Lin 
Zhou$^{[1,4]}$} 
\address{
$^{[1]}$Department of Physics and  Center for Nonlinear Studies\\
Hong Kong Baptist University,  Hong Kong, China \\
$^{[2]}$ Department of Physics, University of Houston, Houston Texas 77204\\
$^{[3]}$ Institute of Applied Physics and Computational Mathematics\\ 
P.O.Box 8009, 100088 Beijing, China\\
$^{[4]}$ Department of Astronomy, Nanjing University, 210093 
Nanjing, China
}
\date{\today} 
\maketitle

\begin{abstract} 

We study kicked quantum systems by using the squeezed state approach. Taking
the kicked quantum harmonic oscillator as an example, we demonstrate that
chaos in an underlying classical system can be enhanced as well as suppressed
by quantum fluctuations. Three different energy diffusions are observed in
the kicked quantum harmonic oscillator, namely, localization, linear
diffusion, and quadratic diffusion. 

\end{abstract} 

\section{Introduction}

A squeezed state is a generalized coherent state, which has a wide
application in many branches of physics such as quantum optics, high
energy physics, etc.  Recent years have witnessed a growing application of
the squeezed state to the field of nonlinear dynamics and chaos
\cite{ZF90a,ZF90b,ZF93,ZL94,ZF95,Mo95,Tsui91,Tsui92,PS94a,PS94b,PS97,LS97}. 
Since the squeezed state approach is a kind of approximation of quantum
mechanics, it is called a {\em semiquantum approach} by people in this
community. The time evolution of expectation values and fluctuations of
the squeezed state is thus named {\em semiquantum dynamics}, or {\em
squeezed state dynamics}. 
 
The main purpose of the squeezed state approach is to study how quantum 
fluctuations
manifest themselves on classical trajectories. This approach starts directly 
from quantum systems with no reference to classical limit. In fact, it
has been shown \cite{ZF95} that the squeezed state dynamics exits even for
systems without a well-defined classical dynamics.  Generally,
the squeezed state approach simplifies the quantum version, and provides a 
complementary way to the semiclassical method. In some 
special cases (unfortunately, we still do not know under what kind of
conditions), it gives us better results than 
the semiclassical method\cite{ZF90a,ZF90b,ZF93,ZL94,ZF95,Tsui91,Tsui92}. 
Therefore, in addition to the semiclassical approach, the squeezed state 
approach is also a very useful tool to study the problem of 
classical-quantum correspondence. 

The squeezed state approach has proven to be very successful in studying
dynamical systems ranging from integrable to many-body nonintegrable
systems. Among many others, we name just a few examples here.  In
calculating the ground state energy for quantum system with potential
$V(q)=-V_0/\cosh^2\alpha q$, Tsui\cite{Tsui92} discovered that the ground
state energy obtained by the squeezed state approach is much closer to the
exact ground state energy than that obtained from the WKB method. More
recently, Pattanayak and Schieve\cite{PS97} applied this approach to a
classically chaotic system, for which the WKB method completely fails.
They have successfully calculated low-lying eigenenergies which agree
within a few percent with the pure quantum (numerical) results.  In
addition to the ground state and/or lower excited state energies, the
squeezed state approach can provide a way to obtain correct
eigenfunctions.  For instance, in applying the squeezed state approach to
a harmonic oscillator, we can obtain not only the eigenenergies but also the
eigenfunctions (including the ground state) exactly, as we shall see in
Appendix A, whereas the semiclassical method fails to yield the exact
eigenfunctions. As a further example of many-body systems, we would like
to mention that we have used the squeezed state approach to the  
one-dimensional quantum Frenkel-Kontoroval model\cite{hlz97}, which is a
nontrivial many-body, non-integrable system. Our results show that the
squeezed state approach  very nicely captures the feature of the quantum
effect, namely, the standard map which determining the coordinates of
the classical ground state, is renormalized to an effective sawtooth map 
in the
quantum case. The squeezed state results agree well with that of the quantum
Monte Carlo method\cite{hlz97}. 

In this paper, we would like to apply the squeezed state to study generic
behaviors of kicked quantum systems. It is well known that, in the 
development
of quantum chaos, kicked quantum systems play a very important role.
Prototypes of these kicked systems are the kicked rotator and kicked harmonic
oscillator (KHO). They represent two different classes of dynamical
systems. On the one hand, the kicked rotator obeys the
Kolmogorov-Arnold-Moser (KAM) theorem. Classically, as the kick strength
increases, invariant curves gradually break up. When the kick strength
exceeds a critical value of $K_c=0.9716...$, the last invariant curve
disappears, and bounded chaos turns into global chaos, characterized by
unbounded diffusion in the momentum direction\cite{Chirikov79}. Quantum
mechanically, the diffusion follows the classical one only up to a certain
time, after which it is completely suppressed, thus leading to dynamical
localization\cite{CCFI79}. This phenomenon was connected to the 
Anderson localization\cite{GPF84}, and was confirmed
experimentally\cite{Moor95}. 

On the other hand, since a harmonic oscillator is a degenerate system, the
KHO model is out of the framework of the KAM theorem, that is, 
diffusion can
occur along stochastic webs for any small kick strength.  As a matter of
fact, the KHO model is not a toy model, it stems from a real physical
system. It describes a charged particle moving in a magnetic field, and
under the disturbance of a wave packet\cite{Zas92}. It has $1\frac{1}{2}$
degrees of freedom.  Classically, it depends on the ratio between
frequency of the harmonic oscillator and that of the external kicks (see
Eq.(\ref{Freqrat})), this system displays abundant structures in phase
space such as crystal, quasicrystal, and stochastic webs  
\cite{Zas92,CSZ88}. Moreover, since the phase space of the KHO is
unbounded and cannot be reduced to a cylinder as in the case of the kicked
rotator model, numerical investigation of the quantum KHO is much more
difficult than that of the kicked rotator model.  Therefore, contrary to
the kicked rotator, only a few attempts have been made in the quantum KHO
\cite{BRZ91,BR95,SS92,Dana94,Frasca97}. A general picture about the quantum
behavior of the KHO is still lacking. 

In this paper, using the squeezed state approach, we are able to obtain
the  diffusion behavior of 
the quantum KHO model not only numerically but also analytically.  
The paper is organized as follows. In Sec. II we 
give a brief
introduction of the squeezed state approach for purpose of self-containment 
In Sec. III, by using the KHO model, we shall demonstrate that
quantum fluctuations will enhance chaos at a small perturbation regime,
whereas it will suppress chaotic diffusion at a large perturbation regime. In
Sec. VI, we shall study the diffusion and localization phenomena. In
Sec. V, comparisons between the squeezed state results and the very few
available quantum results will be given. We shall conclude our paper with
discussions and remarks in Sec. VI. In Appendix A, we apply this approach
to quantize the harmonic oscillator which exactly 
yields eigenenergies and eigenfunctions. In Appendix B, we outline our 
procedure of the pure quantum computation for KHO in some special cases. 

\section{Squeezed State approach}

The squeezed state approach starts from the
time-dependent variational principle (TDVP) formulation
\begin{equation}
\delta \int dt \langle\Psi(t)|i\hbar\frac{\partial}{\partial t} 
-\hat{H}|\Psi(t)\rangle = 0.
\label{TDVP}
\end{equation}
Variation with respect to  $\langle\Psi(t)|$ and $|\Psi(t)\rangle$ 
gives rise to 
the Schr\"odinger equation, and its complex conjugate (see, e.g. 
Ref. \cite{KS81}), respectively. The true solution may be approximated 
by restricting 
the choice of states to a subspace of the full Hilbert space, and finding 
the path along which the above equation is satisfied within this subspace.
In the squeezed state approach, the squeezed coherent state is chosen as 
$|\Psi(t)\rangle$. In this manner, 
as we shall see below, in addition to the dynamics of centroid of 
wave packet, we will also have equations of motion for the 
fluctuations, i.e., the spread of wave packet. Therefore, this approach 
enables us to 
study the effects of the quantum fluctuations on dynamical behavior.

The squeezed state is defined by the ordinary harmonic oscillator 
displacement operator ${\cal D}(\alpha)$ acting on a squeezed vacuum state 
${\cal S}(\beta)|0\rangle$:
\begin{eqnarray}
|\alpha \beta\rangle &=& {\cal D}(\alpha){\cal S}(\beta)|0\rangle\nonumber\\
{\cal D}(\alpha) &=& \exp\left(\alpha \hat{a}^+ 
-\alpha^*\hat{a}\right), 
\label{scsdef}\\
{\cal S}(\beta)&=&  
\exp\left[\frac{1}{2}(\beta\hat{a}^{+2}-\beta^* \hat{a}^2)\right]\nonumber
\end{eqnarray}
$\hat{a}^+$ and $\hat{a}$ are boson creation and annihilation  operators 
which satisfy the canonical commutation relation $[\hat{a},\hat{a}^+]=1$.
The coherent state is just operator ${\cal D}$ acting on the vacuum state 
$|0\rangle$, 
\begin{equation}
|\alpha\rangle = {\cal D}(\alpha)|0\rangle.
\label{Cost}
\end{equation}
In terms of the number eigenstate $|n\rangle$, the coherent state can be 
written as
\begin{equation}
|\alpha\rangle=\exp{\left(-\frac{1}{2}|\alpha|^2\right)}\sum_{n=0}^{\infty}
\frac{\alpha^n}{\sqrt{n!}}|n\rangle.
\label{CS}
\end{equation}
$\alpha$ is eigenvalue of creation operator, i.e.,
\begin{equation}
\hat{a}|\alpha\rangle=\alpha|\alpha\rangle.
\end{equation}
Denoting,
\begin{equation}
\alpha=|\alpha| e^{i\eta},
\end{equation}
and taking the integral over angle $\eta$ from $0$ to $2\pi$ on both 
sides of Eq. (\ref{CS}), we obtain the number eigenstate expressed 
in terms of the coherent state,
\begin{equation}
|n\rangle=\frac{1}{2\pi}\exp{\left(\frac{1}{2}|\alpha|^2\right)}
|\alpha|^{-n}\sqrt{n!}\int_0^{2\pi}d\eta e^{-in\eta}|\alpha\rangle.
\label{Numberst}
\end{equation}
The coordinate and momentum operators are defined as
\begin{eqnarray}
\hat{p}&=& i\sqrt{\frac{\hbar}{2}}(\hat{a}^+-\hat{a}),
\nonumber\\
\hat{q}&=&\sqrt{\frac{\hbar}{2}}
(\hat{a}^+ 
+\hat{a}). 
\end{eqnarray}
Thus we have expectation values and variances 
\begin{eqnarray}
p &\equiv & \langle \Psi(t)|\hat{p}|\Psi(t)\rangle = 
i\sqrt{\frac{\hbar}{2}}(\alpha^* -\alpha),\nonumber\\
q &\equiv &\langle \Psi(t)|\hat{q}|\Psi(t)|\rangle=
\sqrt{\frac{\hbar}{2}}(\alpha^* + \alpha).
\label{avepq}
\end{eqnarray}
\begin{eqnarray}
\Delta q^2 &\equiv &\langle \Psi(t)|(\hat{q} - q)^2|\Psi(t)|\rangle = \hbar 
G,\nonumber\\
\Delta p^2 & \equiv & \langle\Psi(t)|(\hat{p} -p)^2|\Psi(t)|\rangle 
= \hbar\left(\frac{1}{4G} + 4\Pi^2 G\right).
\label{avegpi}
\end{eqnarray}
The canonical coordinates $(G,\Pi)$ were introduced by Jakiw and 
Kerman\cite{JK79} for the quantum fluctuations, and its relation with 
$\beta$ in Eq. (\ref{scsdef}) is\cite{ZF95,Tsui92}
\begin{eqnarray}
G &\equiv &\frac{1}{2}|\cosh|\beta| +\frac{\beta}{|\beta|}\sinh|\beta|
|^2,\nonumber\\
\Pi &\equiv& \frac{i}{2}\frac{\beta^*-\beta}{|\beta|}\frac{\sinh|\beta| 
\cosh|\beta|} {|\cosh|\beta| + \frac{\beta}{|\beta|}\sinh|\beta||^2}.
\label{GPIB}
\end{eqnarray}
In the framwork of the squeezed state, 
the Heisenberg uncertainty relation becomes
\begin{equation}
\Delta q\Delta p =\frac{\hbar}{2}\sqrt{1+16G^2\Pi^2} \ge \frac{\hbar}{2}.
\label{Uncert}
\end{equation}
In fact, the squeezed state $|\alpha\beta\rangle$ is equivalent to the 
Gaussian-type state\cite{Tsui91}
\begin{eqnarray}
|\Psi(t)\rangle 
&\equiv &\frac{1}{(2G)^{1/4}}
\exp\left(\frac{i}{\hbar}(p\hat{q}-q\hat{p})\right)
\exp\left(\frac{1}{2\hbar}\Omega\hat{q}^2\right)|0\rangle\nonumber\\
&=&e^{-i\psi}|\alpha\beta\rangle,
\label{Gaussian}
\end{eqnarray}
where 
\begin{eqnarray}
\Omega &=& 1-\frac{1}{2G} + 2i\Pi,\nonumber\\
e^{-2i\psi} &=& \frac{1}{\sqrt{G}}\left(\cosh|\beta| + 
\frac{\beta}{|\beta|}\sinh|\beta|\right).
\end{eqnarray}
From the TDVP, we obtain
dynamical equations for expectation values and quantum 
fluctuations, 
\begin{equation}
\dot{q} = \frac{\partial H}{\partial p},\quad\dot{p}=-\frac{\partial 
H}{\partial q},
\label{eqmot1}
\end{equation}
\begin{equation}
\hbar \dot{G}=\frac{\partial H}{\partial \Pi}, \quad\hbar\dot{\Pi} 
= - \frac{\partial H}{\partial G},
\label{eqmot2}
\end{equation}
which are canonical equations of motion. 
The dot denotes the time derivative and the Hamiltonian function
$H$ is given by definition
\begin{equation}
H \equiv \langle\Psi(t)|\hat{H}|\Psi(t)\rangle.
\end{equation}
These equations (\ref{eqmot1}), and (\ref{eqmot2}) give us a simple and 
clear picture 
about the motion of the expectation values as well as the evolution of the 
quantum fluctuations, which are responsible for quantum diffusion. 
If the Hamiltonian consists of separate kinetic and potential terms 
such as
\begin{equation}
\hat{H} = \frac{1}{2}\hat{p}^2 + V(\hat{q}),
\label{Hamilt}
\end{equation}
then the Hamiltonian function can be written as
\begin{equation}
H= \frac{1}{2}p^2 + V(q) + \hbar\left(\frac{1}{8G}+2G\Pi^2\right) 
+ \left\{\exp\left[\frac{\hbar}{2}G\left(\frac{\partial}{\partial 
q}\right)^2\right] -1\right\}V(q).
\label{Hamileq}
\end{equation}
In the limit of $\hbar =0$, this Hamiltonian function reduces to the 
classical Hamiltonian.

\begin{center}
{\it Initial conditions}
\end{center}

In order to solve the equations of motion (\ref{eqmot1}) and 
(\ref{eqmot2}),
appropriate initial condition for variables $(q,p)$ and $(G,\Pi)$ 
should be posed. In principle, the initial condition must be physically 
meaningful. Thus the following two conditions 
are generally selected.

(1) {\it Minimum uncertainty.}
The initial state $|\Psi(t_0)\rangle$ should satisfy the condition of 
minimum uncertainty. 
Because $G_0$ is always larger than zero, from the uncertainty principle 
[Eq. (\ref{Uncert})], we have
\begin{equation}
\Pi(t_0)=0.
\end{equation}

(2) {\it Least quantum effect.} We need to determine the initial value of 
$G$. This can be achieved by requesting minimization of $H$ with 
respect to $G$, i.e. 
\begin{equation}
\frac{\partial H}{\partial G} =0,\quad \frac{\partial^2 
H}{\partial G^2} > 0.
\end{equation}
For instance, for a harmonic oscillator (Appendix A) we have
$G(t_0)=1/2\omega_0$.

\section{Enhancement and suppression of chaos}

In this section, we would like to discuss the effect of quantum 
fluctuations on classical chaos. It is commonly argued that quantum 
fluctuations 
suppress classical chaos due to interference, 
while enhancing chaos due to tunneling. This is a rather qualitative 
argument. With the help of the squeezed state 
approach, we are able to do a quantitative analysis of 
quantum fluctuations.  

Up to now,
there have been only a few works on this issue. In  
some models, suppression of chaos comes about, while in others 
enhancement takes place. 
Zhang and co-workers observed suppression 
of chaos in kicked spin and the kicked rotator 
systems\cite{ZF90a,ZF90b,ZL94}. 
Using a one-dimensional problem with a Duffing potential 
without any external perturbation, so that both classical and  
quantum behaviors are regular, Pattanayak and Schieve 
\cite{PS94a} demonstrated that the squeezed state behavior is 
chaotic, and concluded that the quantum fluctuations induce chaos. 
The effect of enhancement was also confirmed in a kicked double 
well model\cite{Mo95}.
We shall see later that the KHO model provides a prototype for studying 
these two effects. The enhancement and suppression can be 
observed by increasing the strength of the kicks.

The Hamiltonian of the quantum KHO model can be written as\cite{Zas92} 
\begin{equation}
\hat{H} =\frac{\hat{p}^2}{2} +
 \frac{\omega_0^2}{2}\hat{q}^2 +
V(\hat{q})\delta_T,
\label{Ham2}
\end{equation}
where 
\begin{equation}
\delta_T=\sum_{n=-\infty}^{\infty} \delta(t-nT).
\label{Delta}
\end{equation}
Using the squeezed state as a trial wave function of 
Hamiltonian (\ref{Ham2}), one can readily obtain
\begin{equation}
H =H_o + H_f + H_p,
\end{equation}
where 
\begin{equation}
H_o = \frac{p^2}{2} + \frac{\omega^2_0 q^2}{2}
\end{equation}
is the 
Hamiltonian of the harmonic oscillator; 
\begin{equation}
H_f = \frac{\Delta p^2}{2} + \frac{\omega_0^2
\Delta q^2}{2}
\end{equation}
denotes the contribution from  quantum 
fluctuations and 
\begin{equation}
H_p =\exp{\left[\frac{\Delta q^2}{2}\left(\frac
{\partial}{\partial q}\right)^2\right]}
V(q) \delta_T
\end{equation}
denots the contribution from the external perturbative  potential. The 
external 
potential can be even or odd. Without loss of generality, we denote it as
\begin{equation}
V(q) = K\Theta(q),\qquad \Theta(q) =\left\{ {
\sin(k_0q),\quad \mbox{for odd} \atop
\cos(k_0q),\quad \mbox{for even.}}
\right.
\end{equation}
From Eqs. (\ref{eqmot1}) and (\ref{eqmot2}), 
we have
\begin{eqnarray}
\dot{q} &=&  p, \nonumber\\
\dot{p}  &=& -\omega_0^2q - K_{eff}{\Theta}'(q) 
\delta_T,\nonumber\\
\dot{G} &=&  4\Pi G,
\label{MOVeq}\\
\dot{\Pi} & =&  \frac{1}{8G^2} -2\Pi^2 -\frac{\omega_0^2}{2}
+ \frac{k_0^2}{2}K_{eff}\Theta(q)\delta_T,\nonumber
\end{eqnarray}
where 
\begin{equation}
K_{eff}= K \exp{\left(-\frac{\hbar k_0^2G}{2}\right)},
\label{Keff}
\end{equation}
is called the effective 
potential, whose physical meaning will be discussed below.
$\Theta'$ is the first derivative of $\Theta$ with respect to $q$.
It is clear to see from Eq. (\ref{MOVeq}) that, in the time interval $ 
nT < t < (n+1)T$, the harmonic oscillator takes the free motion governed 
by
\begin{eqnarray}
\dot{q} &=&  p,\nonumber\\
\dot{p}  &=& -\omega_0^2q, \nonumber\\
\dot{G} &=&  4\Pi G,
\label{FRmeq}\\
\dot{\Pi} & =&  \frac{1}{8G^2} -2\Pi^2 -\frac{\omega_0^2}{2}.\nonumber
\end{eqnarray}
while, at the time $t= (n+1)T$, it is kicked by the
external potential. Therefore, before and after the kick
the momentum and its fluctuation undergo jumps:
\begin{eqnarray}
p_{n+1}(T^+)& =& p_n(T^-) - K 
e^{-\frac{\hbar G_n(T^-)}{2}} \Theta'(q_n(T^-)),\nonumber\\
\Pi_{n+1}(T^+) &=& \Pi_n(T^-) +  
\frac{k_0^2}{2}K e^{-\frac{\hbar G_n(T^-)}{2}}
\Theta(q_n(T^-)).
\label{pPI}
\end{eqnarray}
Eqs. (\ref{FRmeq}) and (\ref{pPI}) are coupled differential equations. 
From these equations, we can readily see that the squeezed state 
dynamics differs from 
the classical one in two ways. (1) There are two additional
dimensions for the squeezed state motion, namely, the particle moves in a 
four dimensional extended phase space. (2) The classical quantities $(q,p)$ 
are coupled with the quantum fluctuations, which make the semiquantal  
motion complicated.
On the one hand, because of these two additional dimensions in phase space, 
we expect that 
invariant curves in classical phase space would not be able to 
prevent the trajectory from penetrating or crossing them semiquantally.
On the other hand, since 
the quantum fluctuations are always positive, they equal zero only in 
the 
limiting case of $\hbar=0$; the effective potential strength $K_{eff}$
is always less than $K$. The reduction of the effective potential acting 
on the wave packet leads to the suppression of chaos.
These two mechanisms coexist in the semiquantal system. They compete 
with each other and determine the dynamical behavior of the underlying 
system. Therefore, we expect that quantum fluctuations may
not only enhance chaos but also suppress 
classical diffusion as well. This argument will be nicely illustrated 
in the following. 

In this section we restrict our calculations on the potential
\begin{equation}
V(q) = -K \sin(q).
\end{equation}
However, we should point 
out that the main conclusions given in this section do not 
depend either on the parity of the potential or the sign of $K$.

\subsection{Enhancement of chaos}

In solving Eqs. (\ref{FRmeq}) and (\ref{pPI}),
we used the seventh and eighth order Runge-Kutta formula with adaptive 
stepsize control. The permissible error is fixed at 
$10^{-12}$. In Fig. 1(a), we plot the classical 
phase space $(q_{cl},p_{cl})$ for a trajectory starting from (0,0) and 
evolving $10^4$ kicks. The parameter $K=0.8$ and  $\sigma =1/\pi$, where
\begin{equation}
\sigma =\frac{\omega_0}{\omega_T}
\label{Freqrat}
\end{equation}
is the ratio between the angular 
frequency of the kicks  $\omega_T$ ($\omega_T=2\pi/T$; $T$ is the period of 
the kicks) 
and the angular frequency of the harmonic oscillator $\omega$. In our 
calculations, we put $\omega_0=1$. It is obvious that, in classical 
phase space, regular (stable islands) and chaotic 
regions coexists. Fig. 1(b) shows the time evolution of expectation 
values $(q,p)$ of the wave packet for $10^4$ kicks.
The wave packet starts from $(q_0,p_0,G_0,\Pi_0) = (0,0,0.5,0)$, 
with $\hbar=0.1$. The selection of initial conditions $G_0=0.5$, and $ 
\Pi_0=0$ follows the minimum 
uncertainty and least quantum effect conditions given in Sec. II. 
Therefore, the initial wave packet has the same width in both $q$ and $p$ 
directions, that is, it is a coherent state. If there are
no kicks, the wave packet starting from this point will evolve exactly along 
the classical particle's trajectories forever. The fluctuations both in 
momentum and coordinate keep constant, 
and are independent of time. In this case, the squeezed state 
dynamics exactly
describes the classical one. Now, if we switch on the kick, the 
situation becomes quite different. As is shown in Fig. 1(a),
the initial point just lies in stochastic 
seas, thus it is evident that 
in the classical case the trajectory will never enter into stable 
islands 
due  to the existence of invariant curves. However, as we predicted 
the invariant curves are not able to prevent the trajectory from 
crossing it via other 
dimensions semiquantally. This is demonstrated by Fig. 1(b), where
all stable islands in the classical phase space are
"visited" by the semiquantal trajectory. 

As a quantitative 
verification, we have 
numerically calculated the maximal Lyapunov exponent
\begin{equation}
 \lambda =\lim_{n\to\infty} \lambda_n
\label{Lyap}
\end{equation}
for the trajectories in both cases.  The 
time behavior of $\lambda_n$ is shown in Fig. 2. It demonstrates
the coexistence of enhancement and suppression. At  
the initial stage the enhancement mechanism is dominant. However, after a 
certain time,  $\lambda_n$ of the squeezed state 
becomes larger 
than its classical counterpart, which means that the enhancement 
mechanism 
becomes dominant, and consequently leads to enhancement of chaos. 
Furthermore, the chaotic motion in the extended 
phase space is characterized by two positive Lyapunov exponents in 
four-dimensional phase space $(q,p,\hbar G,\Pi)$ which could be verified 
readily.

\subsection{Suppression of chaos}

We would like go to another limit, namely, very large 
perturbation, to investigate the suppression of chaos. Classically, 
when $K$  increases, the motion becomes more and more chaotic.
For a sufficient large $K$ such as $K=6$, the classical motion is 
completely chaotic, as shown in Fig. 3(a), where
 $\sigma=1/\pi$.  Like Fig. 1, Fig. 3 is for a 
trajectory starting from the origin and evolving $10^4$ kicks.
The classical chaotic and diffusive process is easily seen from the 
evolution of this phase plot. To demonstrate the 
suppression of chaos (or diffusion process), we start a wave 
packet from $(0,0,0.5,0)$ in the four-dimensional (4D) squeezed state phase 
space. 
The evolution is shown in Fig. 3(b). Comparing Figs. 3(a) and 3(b), it is 
obvious that, in the classical case, the phase space is chaotic and 
diffusive, whereas in the semiquantal case the diffusion process is largely 
slowed down and suppressed. There are invariant-curve-like 
structures that appear in the semiquantal phase space. These structures  
seem to form a barrier for diffusion and thus suppress chaos. 
The suppression of chaos is quantitatively demonstrated by the large
decrease of $\lambda_n$, as is shown in 
Fig. 4, where the suppression mechanism is most important.

To illustrate the suppression, we plot variation of $K_{eff}$ 
with 
time (in units of kicks) in Fig. 5. This plot indeed demonstrates that 
the effective 
perturbation strength is much less than its classical counterpart for 
most of the time during the evolution. This is the reason for the 
suppression.
In fact, the deduction of the effective potential acting on the 
wave packet has a clear physical picture. 
The width of a wave packet centered at 
$(q,p)$ in coordinate space is 
$\Delta q =\sqrt{\hbar G}$, and the external potential has a wavelength
of $2\pi/k_0$. Therefore, there are, in fact, $m(=\sqrt{\hbar 
G}k_0/2\pi)$ periods of external potential acting on the wave packet 
simultaneously. This is quite different from the classical model, where 
only one kick acts on the particle at one time.  
Since the external potential is negative in some places and positive in 
other places, the wider the wave packet, the larger the number of $m$, 
and therefore, the smaller the effective potential acting on the harmonic 
oscillator.
However, if the wave packet $\Delta q$ is so small 
that it is smaller than the period of
the external potential, then the effective potential is large.
As a matter of fact, the effective 
potential $K_{eff}$ in Eq.(\ref{Keff}) can be written as
\begin{equation}
K_{eff} = K \exp{\left(-\frac{m^2}{2\pi^2}\right)}.
\end{equation}

As a significant evidence of the suppression, it is convenient to 
calculate energy diffusion with time $n$ (in units of kicks) for 
an ensemble of trajectories. The diffusion is defined by $ \langle 
E_n\rangle$ subtracts initial averaging energy $\langle E_0 \rangle$,
where 
$\langle\cdots\rangle$ means ensemble the average over many trajectories.
In 
our calculations we 
have taken such an ensemble averaging over $10^4$ initial points 
which are uniformly 
distributed inside a disk area centered at the origin of the phase 
space.
For the classical one, $E_n=\frac{1}{2}(p^2_n+q_n^2)_{cl}$, and  for 
semiquantal dynamics, $ E_n$ is defined by 
\begin{eqnarray}
E_n &=& 
\frac{1}{2}\left\langle \Psi
|\hat{p}^2_n + 
\omega^2_0\hat{q}^2_n|\Psi\right\rangle
\nonumber\\
& = &\frac{1}{2}(p^2_n + \omega^2_0q^2_n) +
\frac{1}{2} \hbar\left(\frac{1}{4G_n} + 4\Pi^2_n G_n
 + \omega^2_0G_n\right) 
\label{SME}
\end{eqnarray}
In Fig. 6, we show the energy diffusion of $K=6$ and $\sigma=1/\pi$ for 
classical 
and semiquantal cases. The suppression of classical diffusion is 
very obvious. 

\subsection{Transition from enhancement to suppression}

We have seen so far that enhancement may happen at the small $K$ regime, and 
suppression at the large $K$ regime. Now we would like to discuss 
the transition from enhancement to 
suppression by changing the strength of the external potential for fixed 
quantum fluctuations. Here we want to show that there exists a 
threshold value of $K_c$ distinguishing enhancement from suppression.

To this end, we need to take an appropriate ensemble 
average over many trajectories in phase space. However, since the 
classical phase space of the KHO model is unbounded, it is impossible to do 
such an 
average over the whole phase space. This makes numerical works very 
difficult. After many numerical experiments, we find a compromise, 
namely, we take the average over a 
disk centered at origin with  radius  $\pi$. We spread $15\times 15$ 
initial points uniformly distributed inside this area. In classical 
case, we calculate the 
Lyapunov exponent for each trajectory after $10^4$ kicks, and plot the 
averaged value denoted as $\langle \lambda_{cl}\rangle$ in Fig. 7. This 
averaged value is in analogy to the 
Kolmogorov entropy in a bounded system. However, strictly speaking, this 
quantity cannot be called Kolmogorov entropy. Nevertheless, this parameter 
captures more or less chaoticity of the underlying system.
In the case of semiquantal dynamics, 
since we have 4D extended phase space, we always have two positive 
Lyapunov exponents. We add these two values, and denote the result as  
$\langle \lambda_{sq}\rangle$. It is
plotted  in Fig. 7 in comparison with the classical result.

From Fig. 7 we can draw the following conclusions:
(1) There exists a certain threshold value of $K_c$. Before this point, 
$\langle \lambda_{sq}\rangle >\langle\lambda_{cl}\rangle$, which means that 
the degree of chaos is 
enhanced; after this point,  $\langle \lambda_{sq}\rangle < 
\langle\lambda_{cl}\rangle$, chaos is suppressed. This critical value 
$K_c$ changes with $\hbar$.
(2) At the region of $K\gg K_c$, $\langle\lambda_{sq}\rangle$ fluctuates 
around a certain value. It does not change with  $K$.
(3) The enhancement and suppression depends largely on $\hbar$.
 
The results discussed in this section are restricted to an irrational
frequency ratio. One might ask, whether our conclusion also applies  
to the
rational frequency ratio. It is well known that the KHO model is a degenerate
system out of the KAM theorem. In classical phase space, there exists a slow
diffusion along the stochastic web for any small value of perturbation. Our
numerical results also show enhancement and suppression. We give one
example of  rational frequency ratio $\sigma=1/4$ and $K=6$ in Fig. 8 
for  suppression. The corresponding Lyapunov exponent is shown in 
Fig. 9. 

Finally, we would like to say a few words about the initial conditions and 
parity of the external
potential.  We have performed a wide range of numerical investigations, and
found that the above discussed qualitative and quantitative conclusions
are independent of the selection of the initial condition and 
the parity of
the external potential. However, the selection of the initial condition must
be physically meaningful, as we discussed in Sec. II. 

Before concluding this section, we would like to discuss 
the connection of suppression to the 
dynamical localization. In fact, this is a challenge to the squeezed state 
approach for this subtle phenomenon. We argued that the dynamical 
localization 
observed in the kicked rotator is due to a suppression of the chaos 
discussed above. In fact, in the limiting case of $\omega_0=0$, 
the KHO model Eq. [\ref{Ham2}] is reduced to the kicked rotator model, in 
which chaotic
diffusion is completely suppressed by the quantum fluctuations and 
results 
in  dynamical localization, a well established fact observed numerically by 
Casati
{\it et al}\cite{CCFI79} almost 20 years ago, and confirmed
recently by experiment\cite{Moor95}.
This was nicely illustrated by Zhang and Lee\cite{ZL94} with the 
squeezed state approach.

\section{Diffusion and localization}

In Sec. III we showed how quantum fluctuations 
enhance and suppress chaos. This fact will definitely affect 
diffusion behavior. For instance, in the limiting  case, when the 
suppression 
becomes dominant localization is expected to happen. In  this 
section we shall give a detailed study of this. In particular, we 
concentrate on the large $K$ regime. This is the most 
difficult region in pure quantum computation. As we shall see the 
squeezed state approach not only provides an easy way to do 
numerical 
calculations but also makes it possible to do some analytical estimations.

The energy $E_n$ of the kicked harmonic oscillator in the squeezed state 
approximation [Eq. (\ref{SME}] can be written as two parts,
\begin{equation}
E_n =  E^c_n+ E^f_n.
\label{SME2}
\end{equation}
$E_n^c$ contains the first two terms in Eq. (\ref{SME}), which is 
due to  the motion of 
the centroid of a wave packet. 
They mimic the effect of classical diffusion (ECD). 
$E_n^f$ includes the last three terms in 
Eq. (\ref{SME}), and is attributed to 
the effects of quantum fluctuations (EQF). These two 
kinds of effects are the main ingredients of the diffusion process in 
the squeezed state dynamics.

The ratio $\sigma$ is an
important quantity, as we shall see soon.
We take $\sigma$, 
the golden mean value $\sigma_g=(\sqrt{5}-1)/2$, and its continued-fraction 
expansion 
$r/s$: $2/3,3/5,5/8$, $\cdots$ as examples. $r$ and $s$ are generated by the 
Fibonacci sequence defined by $F_{0}=1,F_{1}=1$, and $F_{n}=
F_{n-2}+F_{n-1}$ for $n>1$.  Without loss of generality, in all 
calculations, we keep parameters 
$\hbar=1$, $K=6$, $k_0 =1$, and $\omega_0=1$, and the initial point 
is chosen as 
$(0,0,0.5,0)$, which corresponds to the ground state of unperturbed 
quantum harmonic oscillator.

\subsection{Numerical results}

Figures 10 and 11 and 12 and 13 show our numerical 
results of energy diffusion of
\begin{equation}
V(q) = \left\{{
K \cos(k_0q)\atop
K \sin(k_0q)}\right.
\end{equation}
for even and odd parity, respectively. Now we discuss these two cases 
separately.

{\it Even potential}: In this case, 
for all  rational
frequencies, the energy will finally 
goes quadratically with time i.e. $E_{n}\sim n^{2}$. 
As shown in Fig. 10, the slope equals 2 asymptotically in the double 
logarithmic plot. However, for the case $\sigma = r/s$ 
with relatively larger $r$ and $s$, the diffusion starts only after a 
certain time. Before this time the energy 
diffusion is localized. The transient time depends on
the frequency ratios, and is approximately of the order of $(\sigma 
-\sigma_g)^{-1}$. We call this transient region a {\it transient 
dynamical localization} region.

For the irrational case, dynamical localization occurs, as clearly
demonstrated in Fig. 11. This significant phenomenon has been observed and
investigated in various quantum systems in past few years.  

It is worth pointing out that for two trivial cases, i.e. $\sigma = 1/1$ 
and $1/2$, our squeezed state results given here agree {\em completely} with the 
quantum
analytical results of Ref.\cite{BRZ91} which has been the only 
existing analytical results of the quantum diffusion of this 
model 
up to now. This demonstrates the usefulness of the squeezed state approach.
Moreover, by using the squeezed state approach, we have also recovered the 
quantum results obtained numerically by Borgonovi and Rebuzzini \cite{BR95}. 
For more details, see Sec V. 
\\\\
{\it Odd potential}: In this case, quadratic law  is  
observed only in the 
case of rational frequency ratios $\sigma=r/s$ with
odd $s$. For other situations the energy diffuses linearly with 
time approximately, see Figs. 12 and 13.

\subsection{Analytical estimates}

The above numerical results can be understood by analyzing 
evolution equations (\ref{MOVeq}). In fact, we can 
analytically derive the energy diffusion by studying system 
(\ref{MOVeq}).
Starting from Eqs. (\ref{MOVeq}), we find that when the external 
perturbation is 
absent, the two degrees of freedom (DOF's) $(p,q)$ and $(G, \Pi)$ are 
decoupled, and each undergoes free motion.  
In terms of action-angle variable, the Hamiltonian of the free motions 
can be expressed as Eq. (\ref{HOmeq}).
From this formula, we have already seen that
{\em both} free motions of the two DOF's 
are {\em degenerate}. 
It is this degeneracy which makes resonance between the 
two frequencies possible in phase space. Consequently, the 
squeezed state dynamical behavior of the kicked harmonic oscillator is quite 
different 
from that of the kicked rotator\cite{ZL94}. That is, 
the motions of 
the centroid and the fluctuations of the wave packet behave like
an oscillator with fixed frequencies $\omega_0$ and $2\omega_{0}$, 
respectively.

However, when kicks are added, the two degree of freedom becomes  
coupled, and energy may start to diffuse.
It is convenient to express the evolution of system 
in terms of action-angle variables.  From
Eqs. (\ref{IJaction}) and (\ref{HOmeq})
one can readily obtain 4D maps
\begin{eqnarray}
 I_{n+1}&=&I_{n}- K_{eff} \Theta' \left( 
k_0\sqrt{2I_n/\omega_{0}}\sin\phi_{n}
        \right) k_0\sqrt{2I_{n}/\omega_{0}}\cos \phi_{n}, \nonumber\\
  \phi_{n+1}&=&\phi_{n}+ \omega_{0} T +K_{eff}
    \Theta' \left(  k_0\sqrt{2I_{n}/\omega_{0}}\sin \phi_{n}\right)
       k_0/\sqrt{2I_{n}\omega_{0}}\sin \phi_{n},  \nonumber\\
  J_{n+1}&=&J_{n}+K_{eff} \frac{\hbar}{4}k_0^2\sqrt{(4J_{n}+1)^2-1}
     \sin\theta_n \Theta \left( k_0\sqrt{2I_n/\omega_{0}}\sin\phi_{n} 
\right), 
\label{KHOIJ}\\
  \theta_{n+1}&=&\theta_{n}+ 2\omega_{0} T -K_{eff}
    k^{2}\hbar \left( 1-\frac{4J_n+1}{\sqrt{(4J_n+1)^2-1}} \cos \theta_n \right)
    \Theta \left(  k_0\sqrt{2I_{n}/\omega_{0}}\sin \phi_{n}\right). \nonumber
\end{eqnarray}
With this 4D map, we are able to performe analytical estimate of the energy 
diffusion. We shall treat it at two different limiting cases.

\begin{center}
{\it Classical diffusion effect ($\hbar=0$)}
\end{center}

In the classical limit case,  $\hbar=0$, and  $E_n^f= 0$, and the effect of 
classical diffusion
becomes dominant.  Therefore, the change of  energy  during one kick is 
\begin{equation}
\Delta E_n^c= k_0 K \Theta'(k_0 q_n) \left(p_n \cos \omega_0 T-
q_n \omega_0 \sin \omega_0 T\right) 
 + \frac{1}{2} k^2_0K^2 \Theta'^{2}(k_0 q_n).
\label{DeltaI}
\end{equation}
For $K\gg 1$, the orbit can be supposed to be approximately ergodic. After 
ensemble averaging over variables $p$ and $q$, the first two terms 
vanish approximately, thus we 
obtain linear energy diffusion, 
\begin{equation}
E^c_n \sim \langle\Delta E_n^c\rangle n 
\approx \frac{1}{4}k^2_0K^2 n.
\label{Encn}
\end{equation}
Note that the average  of the first two
terms of $\Delta E_n^c$, though is much smaller than 
the last term for large $K$, is nevertheless not exactly zero. This 
results in some oscillations of $E_n$ around the linearity (see Figs. 12 
and 13). 

\begin{center}
{\it Effect of Quantum Fluctuations}
\end{center}
For the second limit case, suppose
a  wave packet starts from $(q,p,G,\Pi) = (0,0,0.5,0)$ 
with an even potential, and the center of the wave packet keeps fixed,
thus $E_n^c \equiv 0$.
The energy diffusion is caused purely by the effects of the quantum 
fluctuations.  In this case we shall analyze the diffusion process for 
two different frequency ratios, i.e., rational and irrational. 
For the rational frequency ratio, let us take the simple case of $\sigma = 
2/3$ as an example. 
During a time of 
3T, there are three kicks acting on the harmonic oscillator. Since the 
frequency of fluctuation is 2$\omega_0$,
$(G,\Pi)$ evolves four periods. Note that the effective amplitude of a kick 
acting on the wave packet is $K_{eff}$ rather than $K$.
Among these three kicks, {\it  only}  that one at a relative small
$G$ affect the free motion of the oscillator significantly. We call
this kick the {\em effective} kick, the effects from other
two kicks can be neglected due to a very large $G$, and consequently a very 
small 
$K_{eff}$. At the time when the next {\em effective} kick is in action,  
$G$ is approximately the same because of the resonance, see Fig. 14. 
Therefore, the increment of $\Pi$ is almost constant,
which means that 
$\Pi_{3n} \approx n$. Thus, from Eq. (\ref{SME}), we obtain 
\begin{equation}
E_n \approx n^2,
\end{equation}
which gives rise to the quadratic law observed
in Figs. 10 and 12.

If $\sigma$ is an irrational number, a very interesting 
thing will happen. From 
Eq. (\ref{Hameff}) we know that the angular variable when a kick is added
is 
\begin{equation}
\theta_n = 2\pi n \sigma + \theta_0 (mod 2\pi).
\end{equation}
This is nothing but 
a pseudorandom number generator, indicating that the jump of $\Pi$ may 
happen 
in upper  $(\Pi > 0)$ and lower parts $(\Pi < 0)$ with the same
probability, and thus the increment of energy in the upper part
will be canceled out by the decrease in the lower part. This leads to 
the localization 
phenomenon observed in Fig. 11. This localization mechanism, resorting to 
a pseudorandom number generator,
reminds us what happens in the kicked rotator, where the localization is
related to Anderson's localization for a quantum particle propagating
in a one-dimensional lattice in the presence of a static-random 
potential \cite{GPF84}. 
Our results imply that it might also be possible to construct a 
connection 
between the kicked harmonic oscillator and the Anderson's problem in the 
framework of the squeezed state approximation. The mechanism  discussed can 
also explain the transient dynamical localization  that
occurs in the case of the rational frequency ratio, as shown in Fig. 
10. Since during the time $t \le (\sigma -\sigma_{g})^{-1}$,
a rational number behaves just like a pseudoirrational number, 
a transient localization phenomenon occurs. 

\begin{center}
{\it General case}
\end{center}

As to the general case of system (\ref{MOVeq}), both the ECD and EQF may 
coexist.  To illustrate this, we consider the case of $\sigma=r/s$, 
where $r$ and $s$
are coprimed integers. Suppose $s$ is odd.  As we have explained above,
between two {\em effective} kicks,  $(q,p)$ and $(G,\Pi)$ evolve 
freely.
Thus the angle variables of $(q,p)$ at the two
successive {\em effective} kicks are $\phi$ and $\phi + 2\pi r$, 
respectively, and
that of $(G,\Pi)$ are $\theta$ and $\theta + 4\pi r$. From Eqs.
(\ref{HOmeq}) and (\ref{MOVeq}), we find that in this case, the 
increment of
$\Delta p$ and $\Delta \Pi$ have the same sign, which means that both the ECD
and EQF are excited, which is independent of the potential parity.
Because the diffusion due to the EQF is $\sim t^2$, which is much faster
than that of the ECD ($\sim t$), thus asymptotically $t^2$
diffusion shows up as shown in Fig. 12. However, if $s$ is 
even, there  is one additional {\em effective} kick between the above 
mentioned two, namely, the $s/2$th kick, at
which the angle variable of $(G,\Pi)$ is $\theta+2\pi r$, while that of
$(q,p)$ is $\phi+ \pi r$. Thus, for the even potential case, the changes of
$\Pi$ due to the two  consecutive {\em effective} kicks have the same sign, 
which
implies that the EQF is excited and $t^2$ diffusion will show up.
For the odd
potential case, the change of $\Pi$ due to the two consecutive {\em 
effective}
kicks have opposite sign, the EQF is thus
suppressed, and we obtain the linear diffusion as seen in Fig. 13. 
 
For the irrational $\sigma$ case, the EQF is suppressed 
and the ECD becomes dominant. If the $(p,q)$ happens 
to be a fixed point in
$(p,q)$ plane as is the case of Fig. 11, localization occurs. This is 
the reason 
for the different diffusion behavior of irrational $\sigma$ in 
Figs. 11 and  13 for even and  odd external potentials, 
respectively.
Please note that $(p,q)=(0,0)$
are the expectation values of all the eigenstates of the harmonic 
oscillator, thus the localization we observed in Fig. 11 is not  
restricted to the case of the ground state, as we discussed up to now, but 
is very general.

\subsection{Transition from localization to diffusion}

The results discussed above focused on diffusion at very large 
perturbation, in this case the underlying classical system is completely 
chaotic. As a further example we would like to demonstrate 
a very interesting and important phenomenon in quantum mechanics, i.e. 
the {\em tunneling effect}. We start a wave packet from point (0,7.5) in the 
classical phase 
space $(q,p)$. Here we have $\sigma=1/5 $ and $V(q) = K \cos q$ with 
$K=0.5$. The wave packet has parameter $G_0=1/2$ and $\Pi_0=0$ according to 
the minimal uncertainty principle. It is a Gaussian wave packet. The 
classical phase space is shown in Fig. 15(a). We see that the starting 
point lies inside a stable island. Classically, a trajectory that starts 
from this point 
will never be able to escape. However, in the quantum case, the situation 
becomes very different. We expect that if the width of the wave packet is 
much smaller than the size of this stable island, the wave packet will be 
confined by this stable island, and thus lead to localization. However, if 
the wave packet becomes 
wider than the size of the stable island, it will spread out. Here we 
demonstrate this quantum phenomenon by the squeezed state approach. 
In Figs. 15(b)-15(d) we plot the energy evolution for different Planck 
constant $\hbar$, which corresponds to different widths of the initial 
wave packet.
At $\hbar=1$ and $2$ we observed a localization phenomenon as in other 
quantum systems. The energy oscillates around a certain value. When 
we increase $\hbar$ further to $\hbar=5$. A transition from 
localization to delocalization occurs which is shown in Fig. 15(d). 

\section{Comparison with quantum results}

To give the reader a clear picture about the accuracy of the squeezed state
approach, we would like to compare our results 
with those obtained from pure quantum computation. However, as mentioned 
above, since diffusion occurs in the whole unbounded phase 
space and cannot be reduced to  motion on a cylinder like the case of 
the kicked rotator, a pure quantum (numerical) investigation is very 
difficult, in particular, in the large $K$ regime. Nevertheless, with a 
large 
amount of CPU time, one would be able to obtain some results in the small 
$K$ regime for short time evolution. This is why there are only limited  
available quantum results in this regime. We will compare them with our 
squeezed state results  here.

First, let us look at the results obtained
in Ref.\cite{BRZ91}, already shown in Fig. 10. As for the case of 1/1 
and 
1/2, our squeezed state results agree completely with the quantum 
one in Ref.\cite{BRZ91}. In this case both squeezed state and pure quantal 
analyses predict $t^2$ growth of the energy. We should mention that 
it was  conjectured in Ref.\cite{BRZ91} that for the general 
case of $1/q$, the energy
growth should less than $t^2$. From our squeezed state analysis we concluded
that there are only three different diffusions: $t^2$, $t$ and
localization. Therefore, our squeezed state analysis also agrees with 
Ref.\cite{BRZ91}'s prediction. 

Now we turn to the results obtained by Borgonovi and Rebuzzini\cite{BR95}.
Since the time unit given in their pictures is not clear, we are not 
able to make any quantitative comparison with our squeezed state results. 
Therefore, we performed quantum calculations by using our own 
program (see Appendix B). All the system parameters are
kept the same as that used by Borgonovi and Rebuzzini.
The results are given in Figs. 16 and 17. These pictures
correspond to different diffusion behaviors. In Fig. 16, a $t^2$
diffusion is obtained ($t$ is in units of kicks); the squeezed state approach
also gives rise to a $t^2$ diffusive behavior, although there is
a difference in prefactor. In Fig. 17, both the quantum and 
squeezed state results show localization around approximately the same 
energy. Please compare these 
two pictures with Figs. 1 and 8 of Borgonovi and Rebuzzini\cite{BR95}, 
respectively.

Our quantum computation technique (Appendix B) is different from that 
used in Ref.\cite{BRZ91}. For a self-consistent test, we have used the same 
parameter as 
that of Fig. 2 in Ref.\cite{BRZ91}, and computed the energy 
diffusion  with our method. We found that our results agree with 
those of Ref.\cite{BRZ91} in every detail.

As already emphasized, because of the unbounded phase space, the quantum 
computation is very time consuming, even for small perturbation. For 
instance,  about ten days CPU time  (IBM RISC System/6000 42T, with 
192 Mbyte RAM) has been spent for Fig. 16, and 20 days CPU time
for Fig. 17.  
 
\section{Conclusions and discussions}

Applying the squeezed state approach to the kicked quantum harmonic
oscillator, we illustrate how the quantum fluctuations affect the
classical dynamics. We have shown that chaoticity can be enhanced as well
as suppressed by the quantum fluctuations. A transition from enhancement
to suppression is observed when we change the strength of the kicks. 

Moreover, with this squeezed state approach, we are able to investigate
the energy diffusion. Three different energy 
diffusion have been observed for the kicked quantum harmonic oscillator, 
namely, localization, linear diffusion, and quadratic diffusion.
The localization is due to strong suppression. 

Though it is a kind of approximation, the squeezed state
can mimic
many true quantum behaviors such as those demonstrated in this
paper. Moreover, there are examples showing that the squeezed state 
approach gives rise to better results than the semiclassical method. However,
under which condition or how far the squeezed state approach can go beyond
the semiclassical method is still an open problem that deserves further 
numerical as well as theoretical study. 

Finally, we would like to remark on the quantization of a quantum 
system whose classical counterpart is chaotic. As is well known, this 
is a tough problem which has attracted tremendous attentions 
in last two decades. Among many others, Gutzwiller's trace formula might 
be the most plausible one\cite{Gutzwiller}. However, this approach 
encountered difficulty 
of divergence, although many important contributions have been 
made to overcome this difficulty. We believe that the squeezed state approach 
might be an alternative way that can contribute to this. Recent 
successful application of this method by Pattanayak and 
Schieve\cite{PS97} to 
calculate eigenenergies of a chaotic system  sheds light on this direction.

\bigskip
\bigskip

{\bf Acknowledgement.}
We would like to thank Dr. L.-H Tang and Dr. F. Borgonovi for many 
stimulating discussions. The work was supported in part by 
grants from the Hong Kong Research Grants Council (RGC) and the Hong 
Kong Baptist University Faculty Research Grant (FRG).

\begin{appendix}
\section{Quantization of the harmonic oscillator by the squeezed 
state approach}

In this appendix, we would like to demonstrate how the squeezed state 
works when applied to a simple quantum system, a harmonic oscillator, 
which is the KHO system with a zero external potential.
The harmonic oscillator is a 
simple but very important model in quantum mechanics. It is an
integrable system, and its eigenenergies as well as eigenfunctions can be
obtained analytically. Therefore, this model is very suitable for testing
approximate methods such as the WKB method and others.

It is well known that the WKB approximation can give us 
exact eigenenergies for this integrable system. However, 
it cannot yields exact eigenfunctions, 
in particular, for the low-lying eigenstates. It gives only 
the envelope of the wave function in the semiclassical limit $\hbar\to 0$. 
In this appendix, we shall demonstrate that when applying the squeezed state 
approach to the harmonic oscillator model, one can obtain  
the energy levels precisely as well as the eigenfunctions.

The harmonic oscillator has 
the Hamiltonian operator
\begin{equation}
\hat{H} = \frac{\hat{p}^2}{2} + \frac{\omega^2_0\hat{q}^2}{2}.
\label{HOH}
\end{equation}
Applying the squeezed state approach to this system, one can easily obtain
\begin{equation}
H= \frac{p^2}{2} + \frac{\omega^2_0q^2}{2} + 
\frac{\hbar}{2}\left(\frac{1}{4G}+4\Pi^2 G + \omega_0^2G\right).
\label{HOHave}
\end{equation}
This Hamiltonian can be expressed in terms of action-angle 
variables
\begin{equation}
H =\omega_o I+ 2\omega_0(J+\hbar\frac{1}{4}),
\label{Hameff}
\end{equation}
where
\begin{equation}
I=\frac{1}{2\pi}\oint p~dq,\quad
J=\frac{1}{2\pi}\oint \Pi d(\hbar G).
\label{IJaction}
\end{equation}
The transformations  between $(q, p)$, $(G,\Pi)$
and  $(I,\phi)$, $(J, \theta) $ have the following forms: 
\begin{eqnarray}
q&=&\sqrt{2I/\omega_o} \sin \phi,\nonumber\\
p&=&\sqrt{2I\omega_o} \cos \phi,\nonumber\\
G&=&\frac{1}{\omega_0}\left((2J+\frac{1}{2})-\sqrt{2J(2J+1)}\cos\theta\right), 
\label{HOmeq}\\
\Pi &=&
\frac{\frac{\omega_0}{2}\sqrt{2J(2J+1)}\sin\theta}{
(2J+\frac{1}{2})-\sqrt{2J(2J+1)}\cos\theta}.\nonumber
\end{eqnarray}
From Eq. (\ref{HOmeq}) one can see that
the motion of {\em both} degrees of freedom  
are {\em degenerate}, namely, $\frac{\partial H_o}{\partial I}$  is 
independent of $I$ and $\frac{\partial H_f}{\partial J}$ independent of 
$J$. 
Furthermore, $(G,\Pi)$ are decoupled from $(q,p)$. 
Thus the centroid of the wave packet goes exactly along the classical 
trajectory. While the fluctuations in momentum and position are 
time independent, i.e., the width of the wave packet keeps constant. 
From the minimal uncertainty principle for the initial condition 
mentioned in Sec. II, we have $G=\frac{1}{2\omega_0}$ and $\Pi=0$. 
Therefore, 
the wave packet along the periodic orbit always keeps its form as a 
coherent state. 

The time evolution of both $(q,p)$ and $(G,\Pi)$ are periodic with 
period $T_0(=2\pi/\omega_0)$ and $T_0/2$, respectively. So we can
apply the EBK quantization to the extended phase space, 
\begin{equation} I = n\hbar,\quad J= m\hbar,\quad n,m=0,1,2,\cdots, 
\end{equation} 
Substituting 
$I$ and $J$ into Eq. (\ref{Hameff}) and keeping in mind that
$G=\frac{1}{2\omega_0}$ and $\Pi=0$, and
thus $m$=0, we obtain the squeezed state eigenenergy, 
\begin{equation} 
E_n=\hbar\omega_0(n+\frac{1}{2}). 
\label{HOEn}
\end{equation} 
This is exactly 
the eigenenergy of the harmonic oscillator. 
Here the zero point energy $\frac{1}{2}\hbar\omega_0$ 
comes into
the formula in a very natural and straightforward way. This is quite 
different
from the WKB method.  In the  WKB method, $\frac{1}{2}\hbar\omega_0$ 
comes from the Maslov phase correction which is necessary because of the 
singularity of the wave function.
In the squeezed state
approach, since we do not have any singularities, the Maslov-Morse
correction is incorporated by the extended variables $G$ and $\Pi$. 
Furthermore, the
energy of the system is in the form of the expectation value of the
underlying Hamiltonian operator, whereas in the usual WKB method, the
energy is taken to be the classical form, i.e., $E=H_{cl}$. This is one of 
the reasons  that people call this approach the {\it semiquantum approach}.

Let us now construct the eigenfunction by the squeezed state approach.
We see that when the trial wave function $|\Psi(t)\rangle$ is transformed to
\begin{equation} 
|\tilde{\Psi}(t)\rangle=\exp\left(\frac{i\lambda(t)}{\hbar}\right)
|\Psi(t)\rangle,
\end{equation} 
the derived variational equations of motion remain invariant. 
Substituting $|\tilde{\Psi}(t)\rangle$ into the Schr\"odinger equation, we 
can determine the
equation of $\lambda(t)$,
\begin{equation}
\lambda(t) = \int^t_{0}dt'\langle\Psi(t')|ih\frac{\partial}{\partial 
t'}-\hat{H}|\Psi(t')\rangle = \lambda_G + \lambda_D.
\label{Eqlambda}
\end{equation}
The second part of the integral corresponds to the dynamical 
phase. We denote it $\lambda_D$.
The first term is the geometrical phase noted as $\lambda_G$, which is 
\begin{equation}
\lambda_G = \frac{1}{2} \int_0^t(p\dot{q}-q\dot{p})dt + \hbar\int_0^t
\dot{\Pi}Gdt.
\label{Geophase}
\end{equation}
Since the motions of $(p,q)$ and $(G,\Pi)$ are periodic, this 
geometrical phase is the Aharanov-Anandan form of Berry's phase. During 
the evolution, each points along the periodic orbit acquires a phase 
factor. However, the dynamical phase does not change
during the evolution, only the geometrical 
phase matters. So the eigenfunction is a weighted sum over points 
of the commensurate periodic orbit\cite{PS94b}. The weight factor at each 
point is an appropriate geometrical phase.
Furthermore, as mentioned above according to the 
requirement of the initial condition, the initial wave packet is 
a coherent wave packet, and it 
does not change its form when cycling along the periodic orbit. 
Substituting expressions for $p$ and $q$ in Eq. (\ref{HOmeq}) into 
Eq. (\ref{Geophase}), and keeping in mind that $\Pi=0$, we can 
easily evaluate the integral 
and obtain the
geometrical phase at time $t$, which is
\begin{equation}
\lambda_G (t) =n\hbar\phi,
\end{equation}
where $\phi = \pi/2-\eta$. Thus the eigenfunction for the bound state 
having the eigenenergy $E_n$ in Eq. (\ref{HOEn}) is
\begin{equation}
c \int^{2\pi}_0 e^{i\frac{\lambda_G}{\hbar}}|\alpha\rangle d\phi=
C \int^{2\pi}_0 e^{-in\eta}|\alpha\rangle d\eta,
\label{HOWF}
\end{equation}
where $|\alpha\rangle$ is the coherent state, 
$\phi$ is the angle in the $(\omega_0q,p)$ plane, and
$C$ is the normalization constant. This is nothing but the number 
eigenstate $|n\rangle$ given in Eq. (\ref{Numberst}) except for the 
prefactor. This constant $C$ can be easily calculated by the normalization.
Therefore, in the coordinate representation, the wave function is
\begin{equation}
\Psi_n(q) = C\int_0^{2\pi} d\eta~e^{-in\eta}\langle q|\alpha\rangle   
\end{equation}
This is the exact wave function of the harmonic oscillator.

\section{Procedure of quantum computation}

In this appendix, we describe our procedure of quantum computation.
Since the Hamiltonian is periodic in time, the Floquet theory can be 
applied. The time evolution can be reduced to evolution of the eigenstate
over one driving period,
\begin{equation}
|\Psi(t+T)\rangle = \hat{U}(T)|\Psi(t)\rangle,
\end{equation}
where
\begin{equation}
\hat{U}(T) = \hat{U}_{free}\hat{U}_{kick} 
= \exp{\left(-i\frac{\hat{H}_0 T}{\hbar}\right)}\exp{\left( 
-i\frac{V(\hat{q})}{\hbar}\right)} 
\end{equation}
is the Floquet operator. 

To simulate quantum diffusion in this degenerate system  
(\ref{Ham2}),
the Fourier spectral method is employed. 
The time interval of 
free propagation
is divided into many slices, each having a width of $\Delta$. For each 
slice 
the evolution operator is factored into a product of kinetic and potential
propagator arranged in a symmetric way so that a full potential step is
sandwiched between two half kinetic steps, namely,
\begin{equation}
\exp{\left(-i\frac{\hat{H}_0\Delta}{\hbar}\right)} =
\exp{\left(-i\frac{\hat{p}^2\Delta}{4\hbar}\right)}
\exp{\left(-i\frac{\omega_0^2\hat{q}^2\Delta}{2\hbar}\right)}
\exp{\left(-i\frac{\hat{p}^2\Delta}{4\hbar}\right)} + O(\Delta^3)
\label{Esplit}
\end{equation}
This technique of symmetrically splitting the kinetic propagator reduces
the error introduced by neglecting the commutator between the kinetic and
potential operators. The error is reduced to $O(\Delta^3)$ 
from $O(\Delta^2)$ in a nonsymmetric
splitting. The kinetic propagation is carried out in
momentum space, since in this space the time evolution is simplified as 
multiplication. The potential step is performed in 
coordinate space for the
same reason. The kick step, performed once per period, is also
done in coordinate space. A fast Fourier transform  routine is 
used to transform wave function between these two spaces. 

Since the KHO model is a degenerate system, a wave packet may diffuse 
rapidly, even to
infinity in both coordinate and momentum space. The average energy of the
wave packet may reach a rather high value during the diffusion. This 
amplifies
the error caused by the approximation made in Eq. (\ref{Esplit}). Therefore, 
the self-adaptative procedure is used to adjust the time slice in
Eq. (\ref{Esplit}) in each period to make sure
that the width of the time slice is much smaller than the inverse energy. 
Second, both coordinate and momentum spaces should be large enough. So a 
large number (32~768) of Fourier components are used in our computations. 
\end{appendix}


\begin{figure}
\caption{
Classical phase space (a) and time evolution of expectation values 
$(p,q)$ of a wave packet (b) 
at $K=0.8$ for an irrational frequency ratio $\sigma=1/\pi$. One classical 
trajectory starts from $(0,0)$. The wave packet starts 
from a 
point having an initial fluctuation parameter $(G_0,\Pi_0)=(0.5,0)$, and 
$\hbar=0.1$. The initial wave packet is a coherent state, i.e. 
it has same width in both $p$ and $q$ direction.}
\end{figure}

\begin{figure}
\caption
{
Time (in units of the kicks) behavior of $\lambda_n$ 
for the trajectory
shown in Fig. 1. The increment of $\lambda_n$ after a certain time in 
the semiquantal case indicates the enhancement of chaos.
} 
\end{figure}

\begin{figure}
\caption{
Same as Fig. 1 but for $K=6$ with an irrational frequency ratio 
$\sigma=1/\pi$
for classical (a) and squeezed state (b) cases ($\hbar=1$). The 
semiquantal  phase space shows an obvious suppression of the classical 
diffusion. }
\end{figure}

\begin{figure}
\caption{
Time behavior of $\lambda_n$
for the trajectory shown in Fig. 3. 
The large decrement of $\lambda_n$ in 
the semiquantal case demonstrates the strong suppression of classical 
diffusion.
} 
\end{figure}

\begin{figure}
\caption{
Time evolution of the effective external potential $K_{eff}$  for 
the orbit shown in Fig. 3.
} 
\end{figure}

\begin{figure}
\caption{Energy
diffusion at $K=6$ for an irrational frequency ratio 
$\sigma=1/\pi$ for classical and semiquantal cases. 
The ensemble averaging is 
taken over $10^4$ initial points which are uniformly distributed in an 
area of the disk centered at (0,0) with a radius of $\pi$ in the phase 
space. 
The diffusion coefficient in the semiquantal case is obviously much smaller 
than that of the classical case, which indicates a strong suppression of 
classical chaos.
} 
\end{figure}

\begin{figure}
\caption{
Transition from enhancement to suppression. The averaged Lyanpunov exponent 
vs the external potential $K$ for classical and semiquantal 
cases with $\hbar=0.1$ and 1. The average is taken over $15\times15$ points 
uniformly distributed inside a disk of radius $\pi$ centered at $(0,0)$. 
Here $\sigma=1/\pi$. 
}
\end{figure}

\begin{figure}
\caption{
Demonstration of suppression for a rational $\sigma=1/4$ frequency ratio 
and 
$K=6$. (a) Classical phase space. (b) Semiquantal $(q_{sq}, p_{sq})$ with 
$\hbar=1$. }
\end{figure}

\begin{figure}
\caption{The Lyapunov exponent for the trajectory shown in Fig. 8.}
\end{figure}

\begin{figure}
\caption{
Evolution of the energy with time (in units of kicks) for 
the case of an even potential with a rational frequency 
ratio. Numbers in the plot indicate the frequency ratio. The case of 1/1 
coincides with that of 1/2.
Note that the curves have slope 2 asymptotically, and a transient dynamical 
localization phenomenon shows up.
}
\end{figure}

\begin{figure}
\caption{
Same as Fig. 10, but for an irrational frequency ratio 
$\sigma=(\sqrt{5}-1)/2$. The localization is obvious.
} 
\end{figure}
 
\begin{figure}
\caption
{Same as Fig. 10, but for the case of odd potential. The dashed 
line with slope 1 is drawn to guide the eye.
} 
\end{figure}

\begin{figure}
\caption{
Same as Fig. 11, but for the case of odd potential. The dashed 
line with slope 1 is drawn to  guide  the eye.} 
\end{figure}

\begin{figure}
\caption{
Evolution $(G,\Pi)$ plot for $K=6$ and $\sigma=2/3$.
}
\end{figure}

\begin{figure}
\caption
{Transition from localization to delocalization of a wave packet 
driving by 
the quantum fluctuations. The wave packet starting from a stable island 
$(q_0,p_0)=(0,7.5)$. $\sigma=1/5$, $\omega_0=1$ and $K=0.5$. (a) Classical 
phase 
space. (b)-(d) Semiquantal energy diffusion. (b) $\hbar=1$. (c) $\hbar=2$. 
(d) $\hbar=5$.}
\end{figure}

\begin{figure}
\caption{
Comparison of semiquantal (dashed line) and quantum (solid line) 
diffusion 
for $\sigma=1/4$, $\omega_0=2/\pi$, $\omega_1=8\pi$, $K=0.5$, and $\hbar=1$. 
The 
trajectory starts from $(q_0,p_0) = (3.15,0)$. The semiquantal trajectory 
has the same $(q_0,p_0)$ and $(G_0,\Pi_0)=(\frac{1}{2\omega_0},0)$.
Please compare it with Fig. 1 in Ref. [21].
}
\end{figure}

\begin{figure}
\caption
{Comparison of semiquantal (top) and quantum (bottom) diffusion 
for $\sigma=(\sqrt{5}-1)/2$, $\omega_0=1$, $K=1$, and $\hbar=1$. 
The trajectory starts from $(q_0,p_0) = (15,0)$. The semiquantal trajectory 
has the same $(q_0,p_0)$ and $(G_0,\Pi_0)=(\frac{1}{2\omega_0},0)$. Please 
compare it with Fig. 8 in Ref. [21].
}
\end{figure}


\begin{references}
\bibitem{ZF90a}
W. M. Zhang, J. M. Yuan, D. H. Feng, R. Qi and J. Tjon, 
Phys. Rev. A {\bf 42}, 3615 (1990).
\bibitem{ZF90b} W. M. Zhang and D. H. Feng, and R. Gilmore, Rev. Mod. 
Phys. {\bf 62}, 867 (1990).
\bibitem{ZF93}
W. M. Zhang and D. H. Feng, Int. J. Mod. Phys. A {\bf 8}, 1417 (1993).
\bibitem{ZL94}
W. M. Zhang and M. T. Lee, Preprint, IP-ASTP-26-94.
\bibitem{ZF95} 
W. M. Zhang and D. H. Feng, Phys. Rep. {\bf 252}, 1 (1995).
\bibitem{Mo95}
S.T. Mo, {\em Master Thesis}, Taiwan, 1995.
\bibitem{Tsui91} 
Y. Tsui and Y. Fujiwara, Prog. Theor. Phys. {\bf 
86}, 443(1991); {\bf 86}, 469 (1991).
\bibitem{Tsui92}
Y.Tsui, Prog. Theor. Phys. {\bf 88}, 911 (1992).
\bibitem{PS94a} 
A. K. Pattanayak and W. C. Schieve,  Phys. Rev. Lett. 
{\bf 72}, 2855 (1994).
\bibitem{PS94b}
A. K. Pattanayak and W. C. Schieve, Phys. Rev. {\bf E 50}, 3601 (1994).
\bibitem{PS97}
A. K. Pattanayak and W. C. Schieve, Phys. Rev. {\bf E 56}, 278 (1997). 
\bibitem{LS97}
W. V. Liu and W. C. Schieve, Phys. Rev. Lett. {\bf 78}, 3278 (1997).
\bibitem{hlz97}
B. Hu, B. Li, and W.-M. Zhang, HKBU-CNS-9702, chao-dyn/9803004.
\bibitem{Chirikov79}
B. V. Chirikov, Phys. Rep. {\bf 52}, 263 (1979).
\bibitem{CCFI79}
G. Casati, B. Chirikov, J. Ford 
and F. M. Izrailev, in {\it Stochastic behavior in Classical and Quantum 
Hamiltonian Systems}, Lecture Notes in Physics Vol. 93, edited by 
G. Casati and J. Ford. (Springer-Verlag, Berlin 1979), P 334
\bibitem{GPF84}
D. R. Grempel, R. E. Prange, and S. Fishman, Phys. Rev. 
{\bf A 29}, 1639 (1984).
\bibitem{Moor95}
F.L. Moore, J. C. Robinson, C. F. Bharucha, B. Sundaram, 
and M. G. Raizen, Phys. Rev. Lett. {\bf 75}, 4598 (1995).
\bibitem{Zas92}
G. M. Zaslavsky, R. Z. Sagdeev, D. A.Usikov and 
A. A. Chernikov, {\it Weak Chaos and Quasiregular Patterns} (Cambridge 
University Press, Cambridge 1992.
\bibitem{CSZ88}
A. A Chernikov, R. Z. Sagdeev, and G. M. Zaslavsky,  Physica D {\bf 33}, 
65 (1988).
\bibitem{BRZ91}
G. P. Berman, V. Yu. Rubaev and G. M. Zaslavsky, 
Nonlinearity {\bf 4}, 543 (1991).
\bibitem{BR95}
F. Borgonovi and L. Rebuzzini, Phys. Rev. {\bf E 52}, 2302 
(1995).
\bibitem{SS92}
D. Shepelyansky and C. Sire, Europhys. Lett. {\bf 20}, 95 (1992).
\bibitem{Dana94}
I. Dana, Phys. Rev. Lett. {\bf 73}, 1609 (1994).
\bibitem{Frasca97}
M. Frasca, Phys. Lett. A {\bf 231}, 344 (1997).
\bibitem{KS81}
P. Kramer and M Saraceno, {\it Geometry of the Time-Dependent 
Variational Principle in Quantum Mechanics}, (Springer-Verlag, Berlin, 
1981).
\bibitem{JK79}
R. Jackiw and A. Kerman, Phys. Lett. {\bf A 71}, 
158 (1979).
\bibitem{Gutzwiller}
M. C. Gutzwiller, {\it Chaos in Classical and Quantum Mechanics}, 
(Springer-Verlag New York 1990).
\end{references}
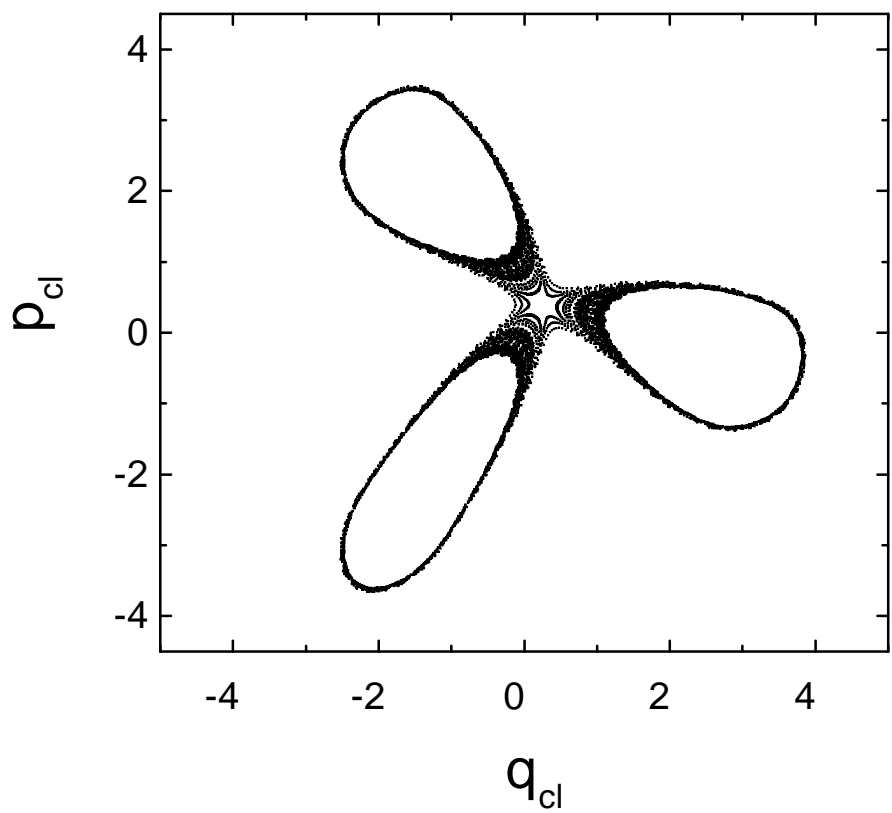
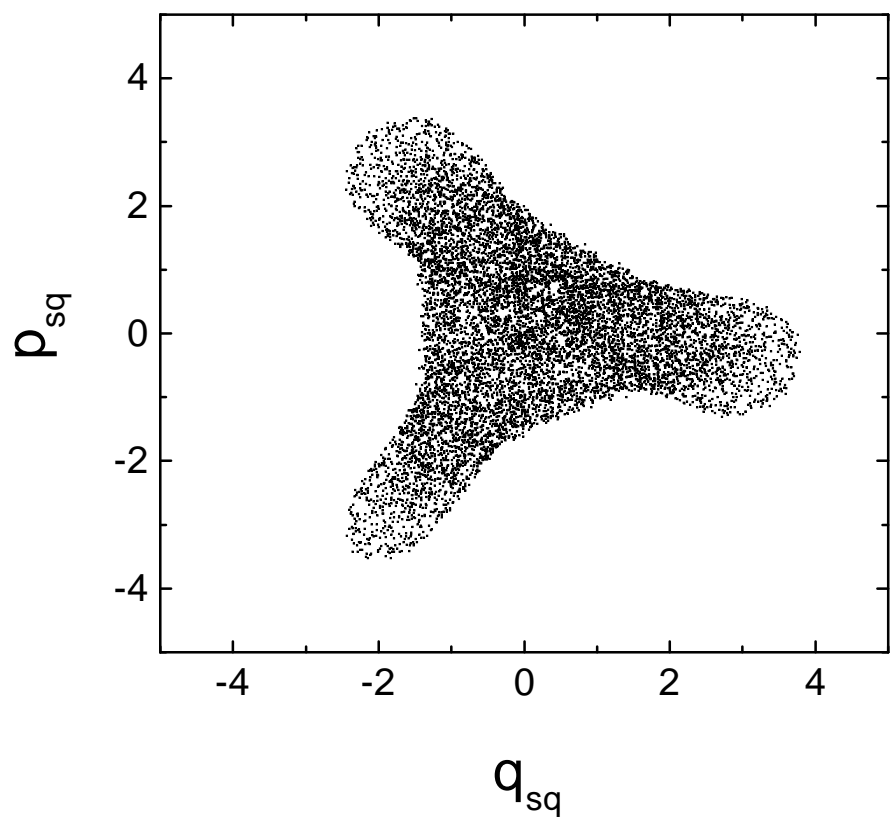
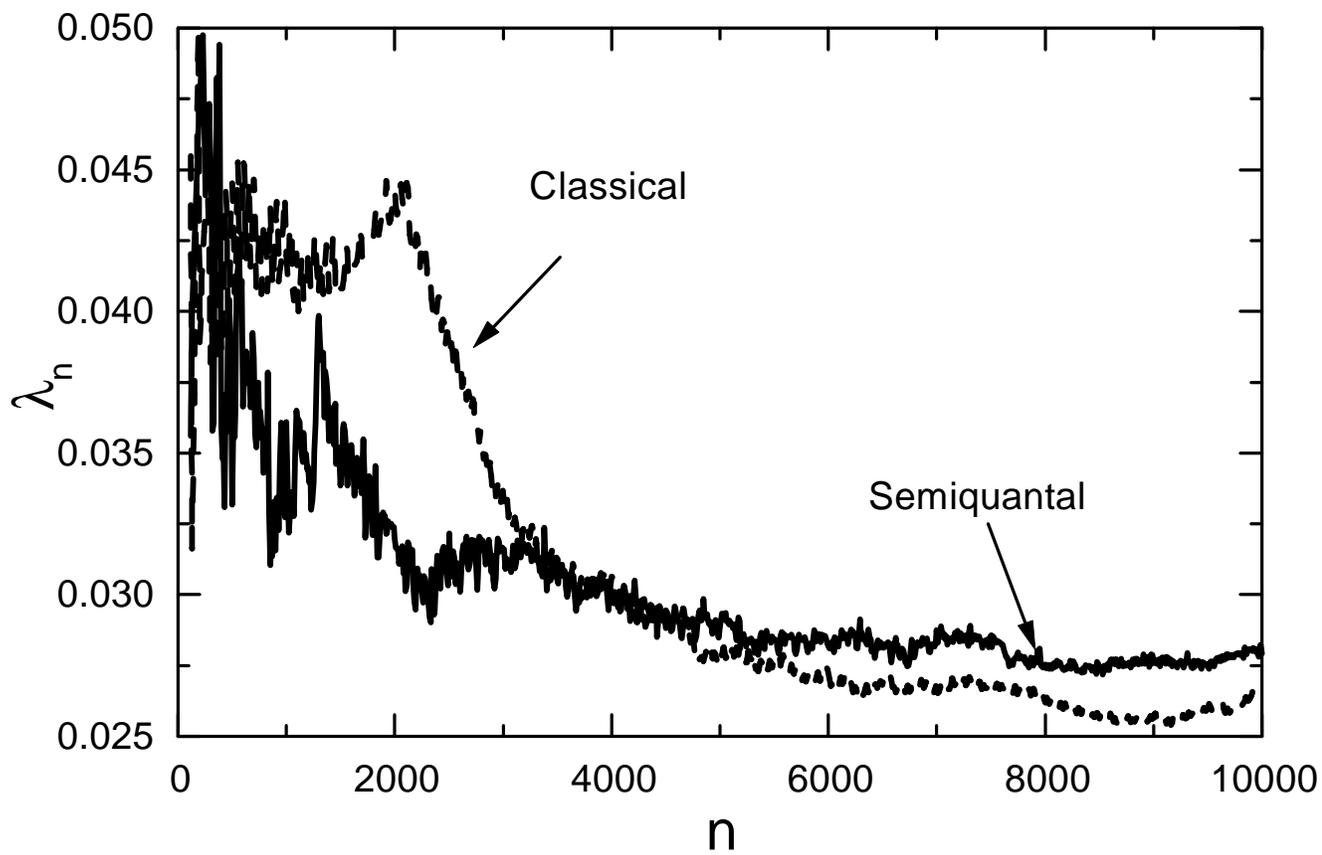
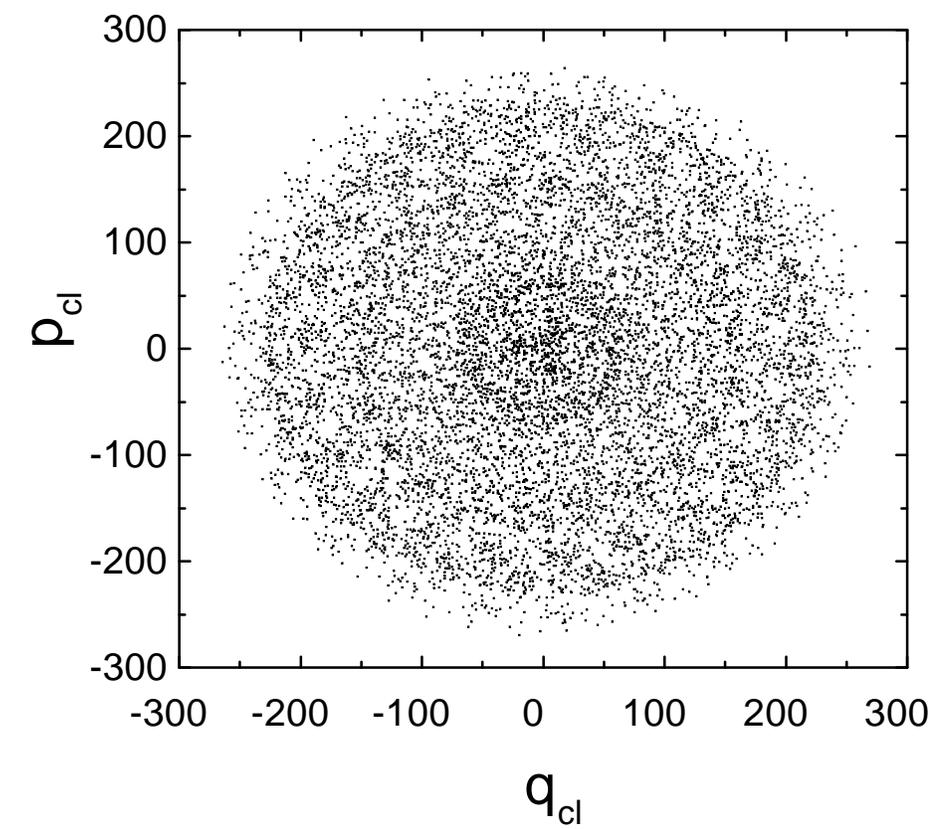
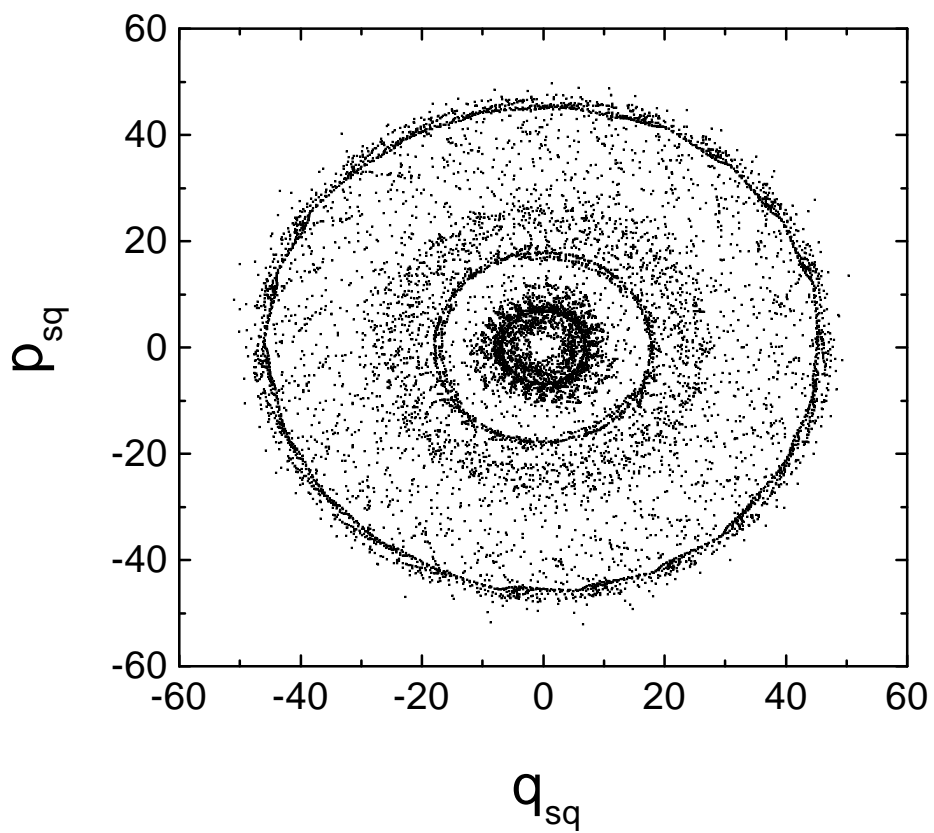
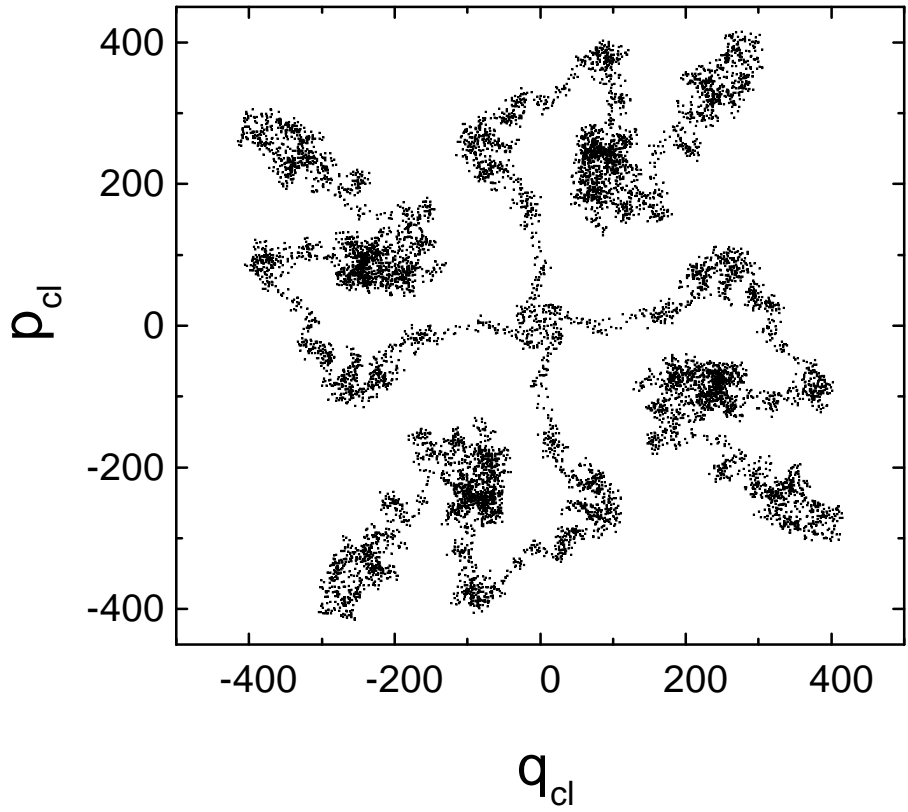
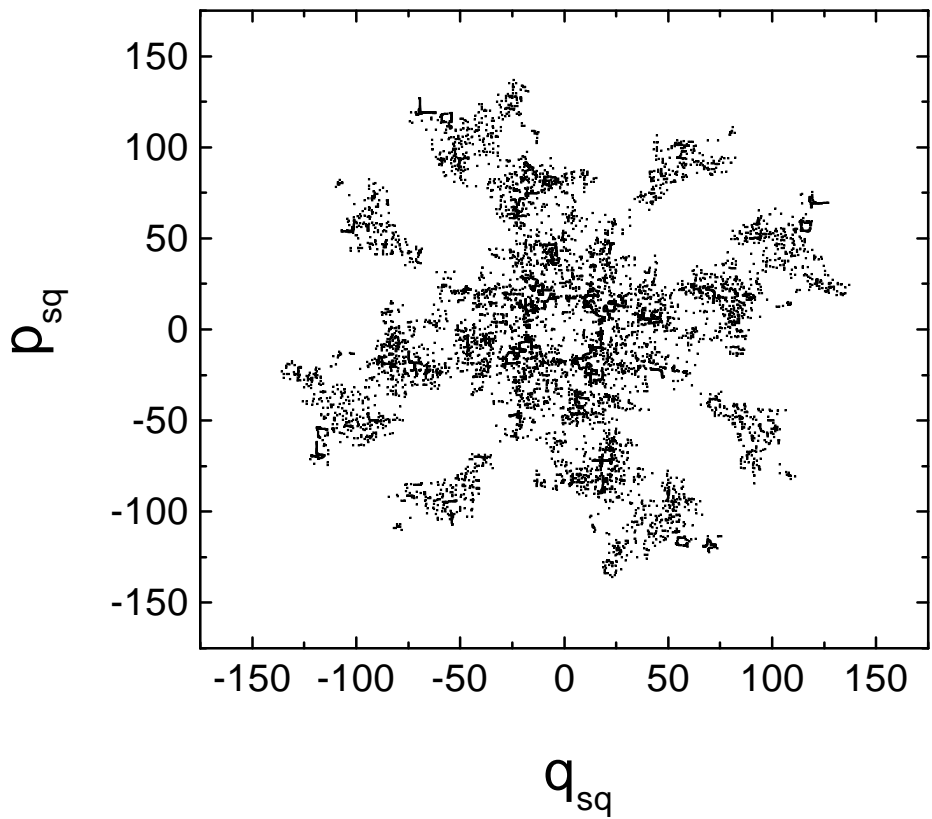
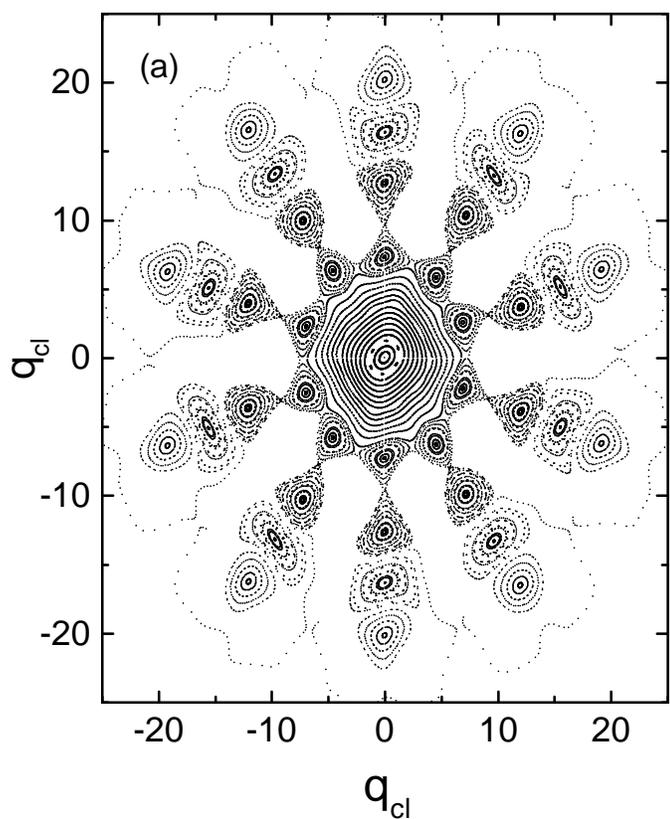
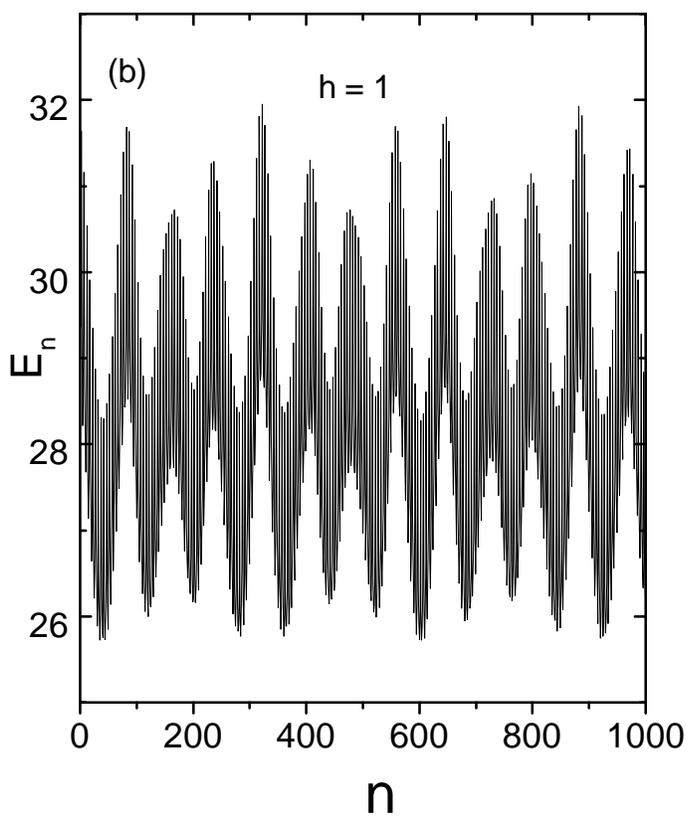
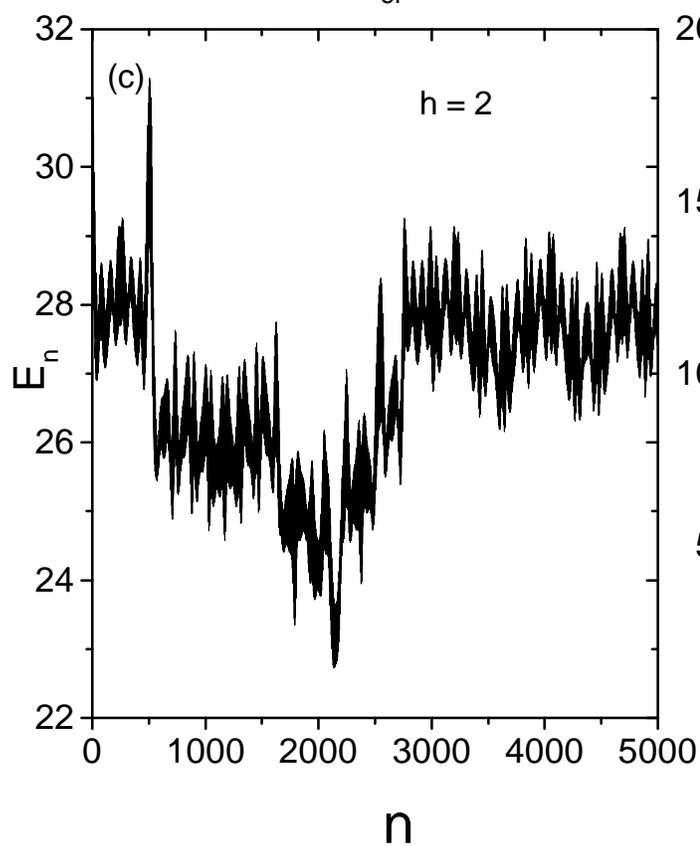
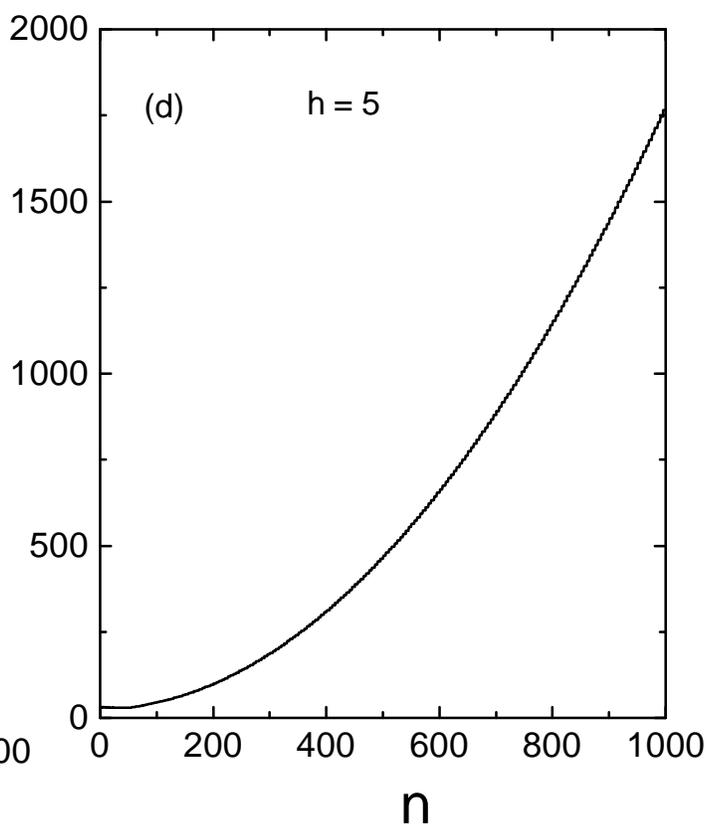
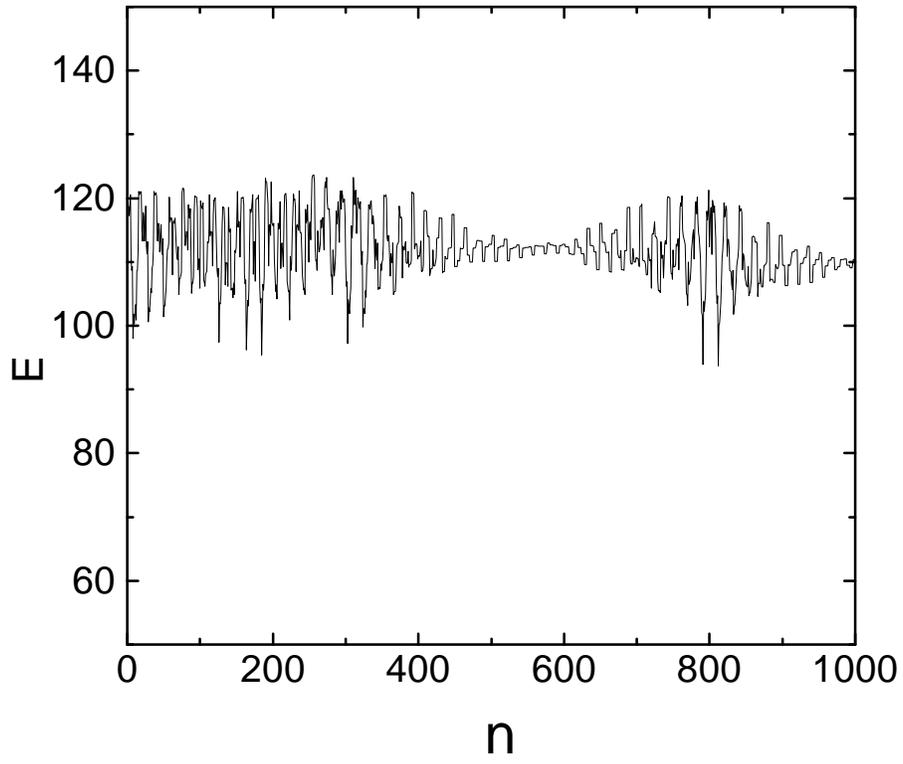
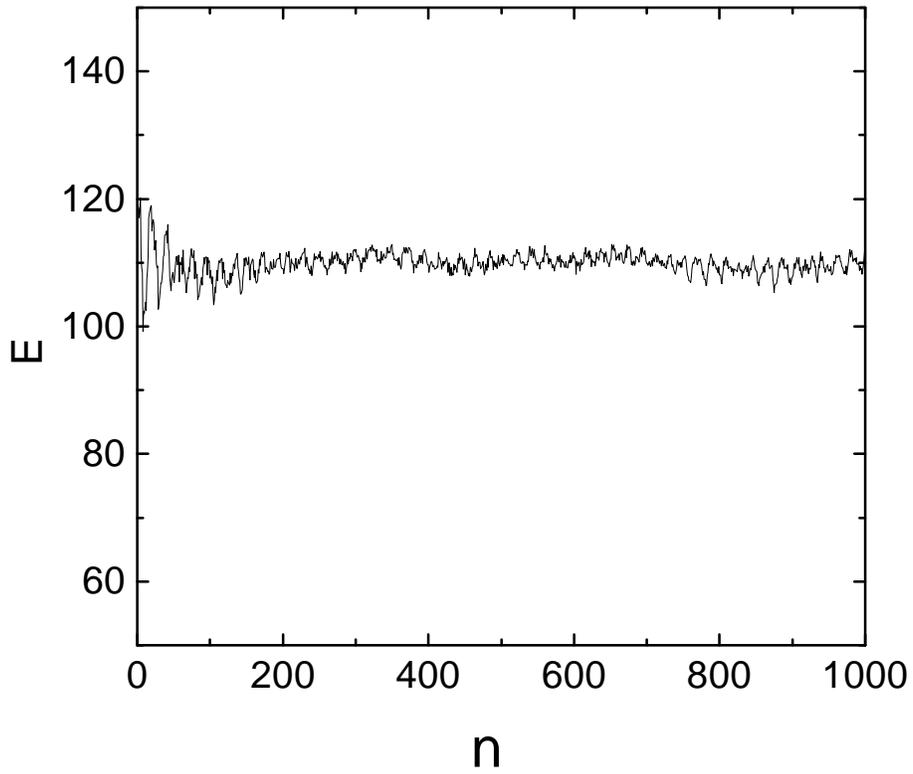
\end{document}